# Effect of PEG binder on gas sensitivity of ferric oxide films synthesized by spin coating and doctor blade methods

P. Samarasekara [1], M.S.P.K. Gunasinghe [1], P.G.D.C.K. Karunarathna [2], H.T.D.S. Madusanka [2] and C.A.N. Fernando [2]

[1]Department of Physics, University of Peradeniya, Peradeniya, Sri Lanka

[2]Department of Nano Science Technology, Wayamba University of Sri Lanka, Kuliyapitiya, Sri Lanka

**Abstract**

Ferric oxide ($\alpha-Fe_2O_3$) thin films were fabricated on amorphous nonconductive and FTO conductive glass substrates starting from iron acetate nanoparticles by the doctor blade and the spin coating techniques. Thin films of $\alpha-Fe_2O_3$ with and without the PEG binder were investigated. Films were subsequently annealed at 500 $^0$C in air for 1 hour to crystallize the phase $\alpha-Fe_2O_3$. The structural and chemical properties of $\alpha-Fe_2O_3$ thin films were determined using XRD and FTIR. According to XRD, the single phase of $\alpha-Fe_2O_3$ crystallized in the thin film after annealing. The optical band gap of synthesized $\alpha-Fe_2O_3$ thin films were determined by UV-Visible spectrums. Gas sensitivity, respond time and recovery time of $\alpha-Fe_2O_3$ thin films were measured in 1000ppm of $CO_2$ gas using AUTO LAB. Thin films synthesized by the doctor blade method were thicker and more uniform than the thin films prepared by the spin coating technique. Gas sensitivity of $\alpha-Fe_2O_3$ films synthesized by the doctor blade method is slightly higher than that of the films fabricated by the spin coating technique. $\alpha-Fe_2O_3$ films with the PEG binder posses a higher gas sensitivity, lower respond time and lower recovery time compared to films without PEG binder. After adding the PEG binder, the optical band gap of $\alpha-Fe_2O_3$ thin films reduces. The reduction of the activation energy is the reason for the improvement of the gas sensitivity after adding the PEG. However, the peak positions of FTIR do not indicate any significant change due to the adding of the PEG. Adding the binder increased the gas sensitivity of iron oxide by 15.44%. Both FTIR and UV-Visible data confirm the formation of $\alpha-Fe_2O_3$ in the thin film.

# 1. Introduction:

Iron oxide is a prime candidate of magnetic data storage, biosensing, drug delivery, magnetic resonant imaging, photodynamic therapy and agriculture applications. $Fe_2O_3$ is a dark red material with band gap 2.2 eV. The resistivity and activation energy of $Fe_2O_3$ at room temperature are $6.5 \times 10^5$ $\Omega$cm and 0.728eV, respectively. Preparation of iron oxide in thin film form has been received a wide attraction due to the potential applications of $Fe_2O_3$. Thin films of iron oxide have been fabricated by both post-oxidation of pure Fe ultra-thin films and by evaporating Fe onto the MO substrates [1]. Colored iron oxide thin films have been synthesized by Sol-gel technique [2]. Iron oxide thin films have been deposited on fused quartz substrate using simple metal organic deposition from Fe-(III) acetylacetonate as the organic precursor [3]. $Fe_3O_4$ thin films have been sputtered using a target consisting of a mixture of $Fe_3O_4$ and $Fe_2O_3$ onto Si and glass substrates [4]. Also thin films of hematite have been synthesized using pulsed laser depositions (PLD) [5]. $Fe_2O_3$ thin film gas sensor sensitive to organic vapors and hydrogen gas have been synthesized using cathodic sputtering [6]. $Fe_2O_3$ gas sensing films have been deposited by normal pressure chemical vapor deposition to detect acetone and alcohol [7]. $Fe_2O_3$ thick film sensors have been used to detect $CH_4$, $H_2$ and $NH_3$ [8]. Hollow balls of nano $Fe_2O_3$ has been employed to detect dimethyl methylphosphonate at room temperature [9]. It is possible to control the gas sensing properties of hematite nanocrystals by controlling the morphology [10]. Gas sensing properties of p-type $\alpha$–$Fe_2O_3$ polyhedral particles have been investigated [11]. Durability and stability of $\alpha$–$Fe_2O_3$ oxide micro and nano structures have been studied [12].

Thin films have been synthesized using techniques incorporated with vacuum by us previously [13, 14, 15, 16]. Compared to the expensive techniques required vacuum, spin coating and doctor blade techniques were found to be low cost and fast. In this manuscript, the band gap determined using an optical method is presented. However, it is possible to determine the band gap using electrical conductivity measurements too [17]. Iron oxide is one of the most famous magnetic materials. Magnetic properties of ferromagnetic and ferrite thin films have been investigated by us using the modified second and third order perturbed Heisenberg Hamiltonian by us [18-22]. Gas sensitivity of many materials has been investigated in many different gases [23-30].

The gas sensitivity, surface morphology, structural, chemical and optical properties of $\alpha$–$Fe_2O_3$ thin film samples fabricated by doctor blade method are presented in this manuscript. Structural, chemical and optical properties were determined by X- ray diffraction (XRD), Fourier Transform Infrared (FTIR), UV-Visible techniques, respectively. The gas sensitivity was calculated by measuring the resistance of the sample in the particular gas and the atmospheric air. Because $CO_2$ gas creates the highest pollution in urban and industrial areas, the gas sensitivity of $\alpha$–$Fe_2O_3$ films in $CO_2$ gas was studied.

**2. Experimental:**

(a) Sample preparation:

First 1.5002 g of iron acetate nanoparticles were dissolved in 10 ml of water to prepare a solution of 2 M. It was stirred on a magnetic stirrer at 600 rpm for 1 hour to mix the solution. The solution was placed inside a furnace at 500 $^0$C for two hours with 10 $^0$C min$^{-1}$ heating rate. Then 0.0503 g of polyethylene glycol (PEG) was mixed with 8 ml of water. It was placed on a magnetic stirrer and stirred at 45 $^0$C temperature for 15 minutes. Prepared PEG solution (2 ml) was added to iron acetate solution, and few drops of ethanol was added to it. Then the solution was placed on the magnetic stirrer, and it was stirred at 600 rpm for two hours at 50 $^0$C temperature. Finally the prepared iron acetate-PEG solution was applied to a conductive FTO glass plate or a normal non-conductive glass plate to prepare $\alpha$–$Fe_2O_3$ thin films using doctor blade and spin coating methods. Samples prepared on non-conductive glass plates were employed for XRD, FTIR and UV-Visible absorption measurements. Thin films grown on conductive FTO glass plates were used for gas sensitivity measurements. Solutions with and without the PEG binder were applied to the glass plates in order to study the effect of PEG on gas sensitivity. FTO glass plates with the area of 3.5 cm x 2 cm were used. An area of 0.6 cm x 2 cm was scratched in the middle of the conducting side on FTO glass by using very smooth glass cutter. Then glass slides were well cleaned using ethanol. If thin films are prepared on unscratched conductive glass plates, then only the conductive glass plates carry the electric current between two electrodes. When the scratched glass slide is used to fabricate the gas sensor, only the $\alpha$–$Fe_2O_3$ layer conducts the electric current.

In the doctor blade method, cello tapes glued to the edges of the glass plates were used to control the thickness. In spin coating method, the samples were synthesized at spin speeds of 300, 400, 500, 600, 700 and 800 rpm for 2 minutes. First the prepared samples were heated on hot plate at 50 $^0$C temperature for 1 hour. It is important to stabilize the PEG binder, because PEG is stable at 45 - 50 $^0$C temperature range. Then the thin films were cooled down in normal air for 2 hours. Thereafter, they were placed inside the oven at 150 $^0$C temperature for 1 hour to remove excess oxygen and water vapor from the sample. Next the thin films were annealed in the furnace at 500 $^0$C for one hour in air to crystallize the phase of $\alpha-Fe_2O_3$. Above 465 $^0$C temperature, iron acetate removes its partially bonded oxygen atoms and converts to stable $\alpha-Fe_2O_3$, and that converted $Fe_2O_3$ is not converted back to the iron acetate when the samples is cooled. This implies that the reaction is irreversible.

(b) Structural and chemical property measurements:

Structure of the thin films was investigated using a X- Ray diffractometer Rigaku Ultima IV with CuK$_\alpha$ radiations. Chemical properties of $\alpha-Fe_2O_3$ thin films were measured by a SHIMADZU IRAffinity-1S Fourier Transform Infra Red spectrometer. The optical energy gap of $\alpha-Fe_2O_3$ thin films was determined by a Shimadzu 1800 UV/Visible spectrometer. The operating wavelengths of the UV/Visible spectrometer were in the range from 190 to 1100 nm.

(c) Gas sensitivity measurements:

The prepared $\alpha-Fe_2O_3$ samples were connected to a 5 V power supply for 6 hours to stabilize the sensor. Thereafter, the gas sensors were used to identify tested gases. Gold coated electrodes and wires were used for all the connections. The sensor electrode wires were connected to a Keithley 6400 source meter unit to measure the current carrying through the gas sensor. Then the Keithley 6400 source meter was adjusted for current measuring mode, and 5.0 V was applied to the gas sensor. Next the measured time period was adjusted to 5000 s, and "AUTO LAB" measuring unit was switched on to measure the current carrying through the gas sensor. Thereafter, a few minutes were given to stabilize the current through the gas sensor. After stabilizing the current, some known amount of $CO_2$ gas (1000 ppm) was injected in to the glass chamber using a syringe. Then electric current increased and reached the saturated value of the current, and this saturated current was noted down. The time taken to reach the saturated current was also measured. This is called the respond time. Thereafter, normal atmospheric air was pumped in to the glass chamber to remove the injected $CO_2$ gas, and the air was pumped continuously in to the glass chamber until the current reading returned to initial stable value. The time taken to reach the initial stable value was also measured. This is called the recovery time. Then the air pump was switched off and, few minutes were given to stabilize the gas sensor. Thereafter, $CO_2$ gas was injected again to the gas chamber, and above procedure was repeated to obtain another current variation cycle. This procedure was repeated for the gas sensors prepared with the binder (PEG) and without the binder, and the current response variation was compared.

## 3. Results and Discussion:

Figure 1 represents the XRD patterns of the $\alpha-Fe_2O_3$ thin films fabricated with and without the PEG binder. According to XRD patterns, single phase of $\alpha-Fe_2O_3$ could be crystallized after annealing the sample at 500 $^0C$ for one hour in air. Adding the PEG binder does not change the structure of the sample. The samples annealed at temperatures below 500 $^0C$ were not crystallized well. The samples annealed above 500 $^0C$ were not stable. Lattice spacing and the Miller indices found from XRD patterns in Figure 1 are tabulated in Table 1.

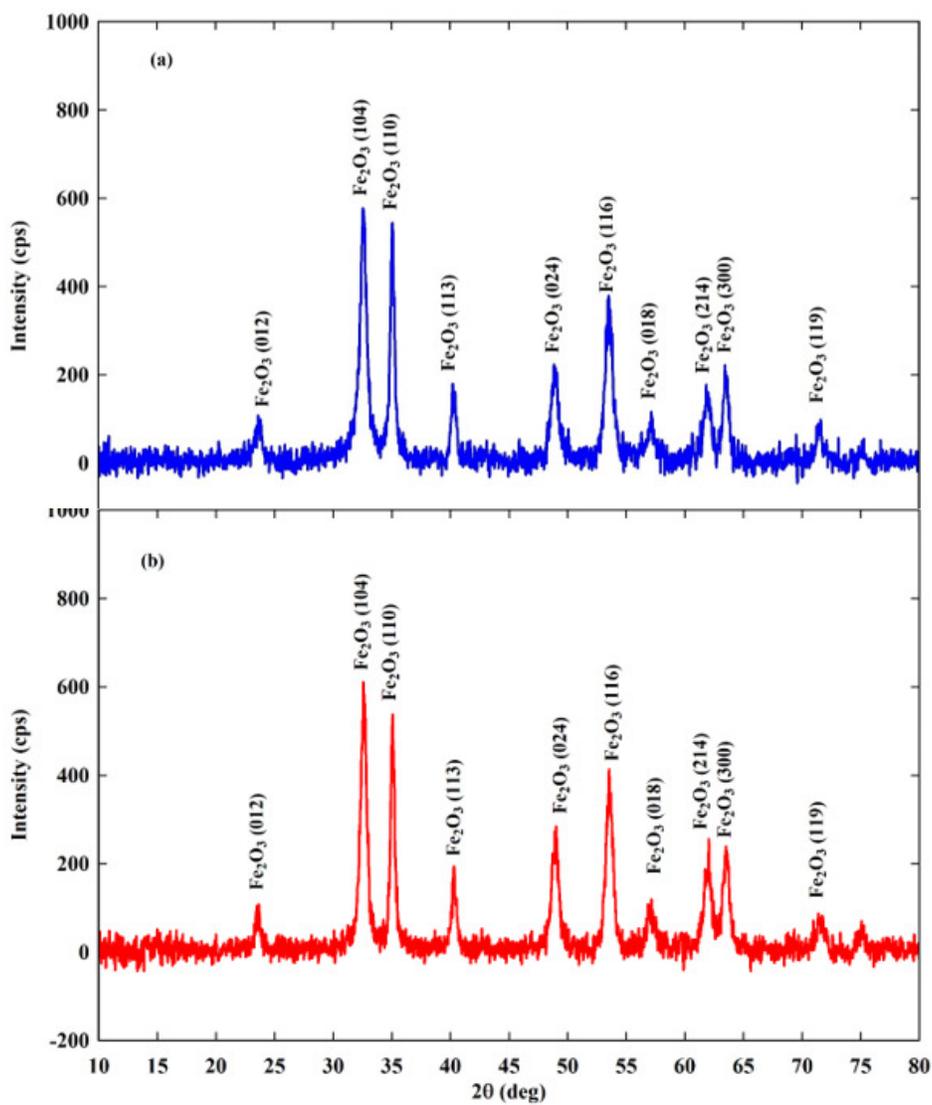

Figure 1: XRD patterns of α–Fe$_2$O$_3$ thin films (a) with the PEG binder (b) without the PEG binder.

| Miller Indices | Diffraction angle ($2\theta$) | Bragg angle ($\theta$) | Lattice spacing (d) ($A^0$) |
|---|---|---|---|
| (012) | $24.16^0$ | $12.08^0$ | 3.681 |
| (104) | $33.18^0$ | $16.59^0$ | 2.698 |
| (110) | $35.65^0$ | $17.33^0$ | 2.586 |
| (113) | $40.88^0$ | $20.44^0$ | 2.206 |
| (024) | $49.50^0$ | $24.75^0$ | 1.840 |
| (116) | $54.17^0$ | $27.09^0$ | 1.691 |
| (018) | $57.57^0$ | $28.78^0$ | 1.600 |
| (214) | $62.47^0$ | $31.24^0$ | 1.485 |
| (300) | $64.01^0$ | $32.01^0$ | 1.453 |
| (119) | $72.53^0$ | $36.27^0$ | 1.302 |

Table 1: Miller indices and lattice spacing found from XRD patterns.

The FTIR spectrums of $\alpha-Fe_2O_3$ thin films with and without the PEG binder are shown in Figure 2. FTIR peaks appear at 456-468 $cm^{-1}$ and 549-560 $cm^{-1}$ [31]. These are the peaks corresponding to Fe-O bonds in $\alpha-Fe_2O_3$. These FTIR data confirm the formation of the $\alpha-Fe_2O_3$ phase in the thin film.

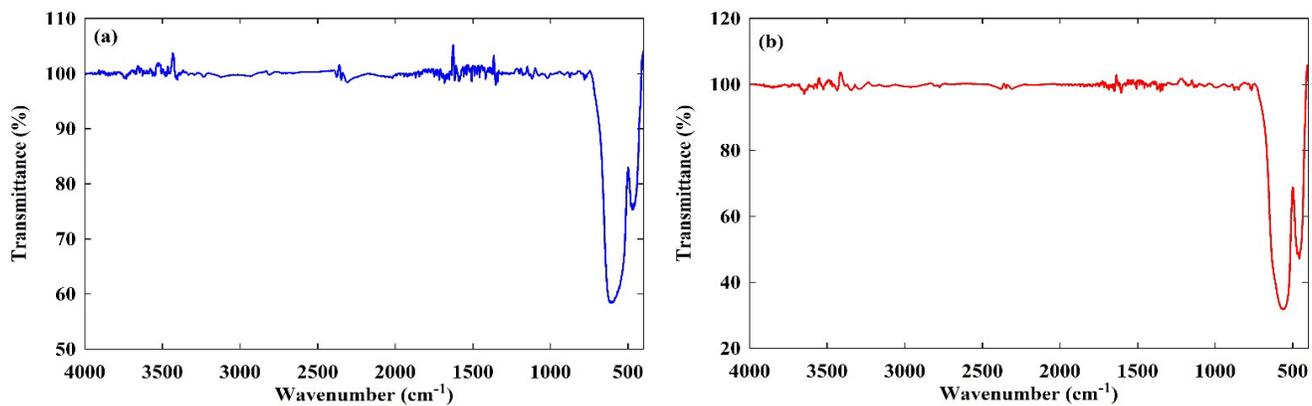

Figure 2: FTIR spectrum of α−Fe$_2$O$_3$ thin films (a) with the binder (b) without the binder.

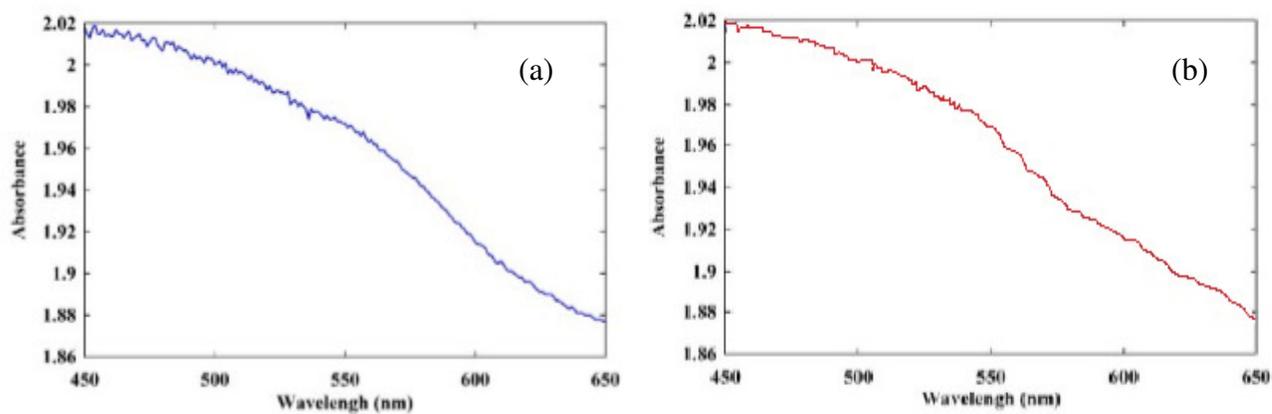

Figure 3: UV-Visible absorption spectrum of α−Fe$_2$O$_3$ thin films (a) with the PEG binder (b) without the PEG binder.

Figure 3 shows the UV-Visible spectrums of α−Fe$_2$O$_3$ thin films with and without the PEG binder. The optical band gap ($E_g$) was found using

$$E_g = \frac{hc}{\lambda} \qquad (1)$$

Where $h$, $c$ and $\lambda$ are the Planck's constant, speed of light and absorption edge, respectively. Absorption edge is the intercept of tangential line drawn to the part of the absorption curve with the highest slop on the horizontal axis. Absorption edges of the samples with and without the binder are 648 and 631 nm, respectively. The optical band gaps of α–$Fe_2O_3$ samples with and without the binder are 1.92 and 1.98 eV, respectively. The optical band gap of the film without the PEG binder is the same as the band gap of pure α–$Fe_2O_3$, which indicates the formation of α–$Fe_2O_3$ in the thin film. Adding a foreign material contributes more energy level to the energy levels of pure α–$Fe_2O_3$. These extra energy levels reduce the effective band gap.

$$\text{Gas sensitivity} = \frac{|R_g - R_a|}{R_a} \times 100\% \qquad (2)$$

Here $R_g$ and $R_a$ are the resistances of the sample in the particular gas and atmospheric air, respectively.

Figure 4 shows the variation of the electric current, the resistance and the gas sensitivity of α–$Fe_2O_3$ thin film samples with and without the PEG binder measured in 1000ppm of $CO_2$ gas at the room temperature. Both these samples were annealed at 500 $^0C$ for 1 hour in air. After applying a voltage of 5.00 V across α–$Fe_2O_3$ layer, the current ($I$) was in the 2-3 mA range. Here red and black curves represent the data of α–$Fe_2O_3$ with and without the PEG binder, respectively. The resistance ($R$) was calculated using

$$R = \frac{5V}{I} \qquad (3)$$

Gas sensitivity of α–$Fe_2O_3$ increased from 51.04% to 66.48% after adding the PEG binder. Response time slightly decreased from 821 to 819 s after adding the binder. In addition, recovery time slightly decreased from 624 to 619 s after adding the PEG binder. Gas sensors with high sensitivity, low respond time and low recovery time indicate best performances. Therefore, all the gas sensing properties can be improved by adding the PEG binder. Adding the binder increased the gas sensitivity of iron oxide by 15.44%. According to the Figure 3, adding the binder reduced the optical band gap of $\alpha-Fe_2O_3$. The reduction of the optical band gap is attributed to the increase of the gas sensitivity. When the optical band gap of the sample is narrow, more electrons can be transferred to the conduction band by supplying a less amount of energy. The activation energy reduces with the decrease of the optical band gap. The electrical conductivity increases with the decrease of the activation energy at one particular temperature according to the Arrhenius equation given below. As a result, the gas sensitivity enhances.

$$\sigma = \sigma_0 \exp(-\frac{E_a}{kT}) \qquad (4)$$

Where $\sigma$, $\sigma_0$, k, T and $E_a$ are the conductivity at temperature T, conductivity at absolute zero, Boltzmann's constant, absolute temperature and the activation energy, respectively.

When the binder PEG was added to the $\alpha-Fe_2O_3$ sample, the surface structure of $\alpha-Fe_2O_3$ will attract more oxygen atoms from the atmosphere. That surface binder oxygen is called partially bindered oxygen in to the semiconductor surface, and they donate electrons in to the conduction band of the $\alpha-Fe_2O_3$. As a result, the free electron concentration in the conduction band of $\alpha-Fe_2O_3$ increases, and that causes to increase electrical current passing through the top surface of the gas sensor. Therefore, the electrical conductivity of $\alpha-Fe_2O_3$ partially increases with the binding material.

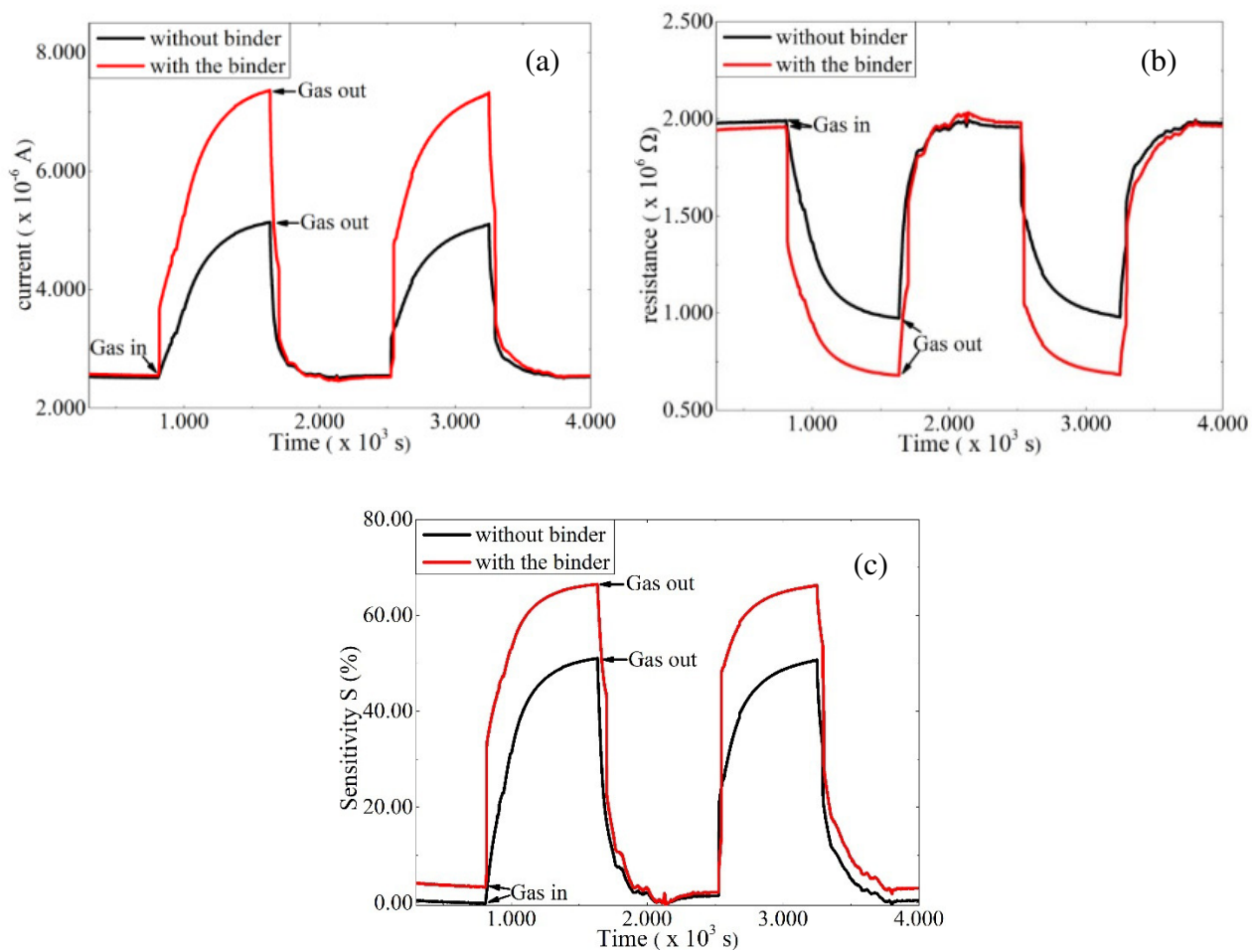

Figure 4: The graphical representations of (a) current (b) resistance and (c) gas sensitivity of α–$Fe_2O_3$ thin films in $CO_2$ gas with and without the PEG binder.

Figure 5 represents the electric current, the resistance and the gas sensitivity of α–$Fe_2O_3$ thin films with the PEG binder prepared by the doctor blade and spin coating techniques. These samples were measured in 1000ppm of $CO_2$ gas at room temperature. Black and red curves represent the α–$Fe_2O_3$ thin films prepared by spin coating and doctor blade techniques, respectively. The samples synthesized using the doctor blade method was found to be more uniform and thicker than the samples prepared by the spin coating technique. The spin coated sample was synthesized at spin speed of 500 rpm for 2 minutes. While the samples prepared at the spin speed of 300 rpm were less uniform, the samples prepared at the spin speed of 800 rpm were thinner. 500 rpm was found to be the best spin speed. Both these thin films were annealed at 500 $^0$C in air for 1 hour. Table 2 shows the gas sensitivity, respond time and recovery time of the thin films fabricated by the doctor blade and spin coating techniques, which were calculated from the graphs in Figure 5. The gas sensitivity of the thin film prepared by the doctor blade method is slightly higher than that of the thin film prepared by the spin coating technique. The reason is attributed to the thickness and the uniformity of the samples synthesized by the doctor blade method. Both the respond and recovery times of the gas sensors deposited by the doctor blade method are higher than those of the samples prepared by the spin coating technique.

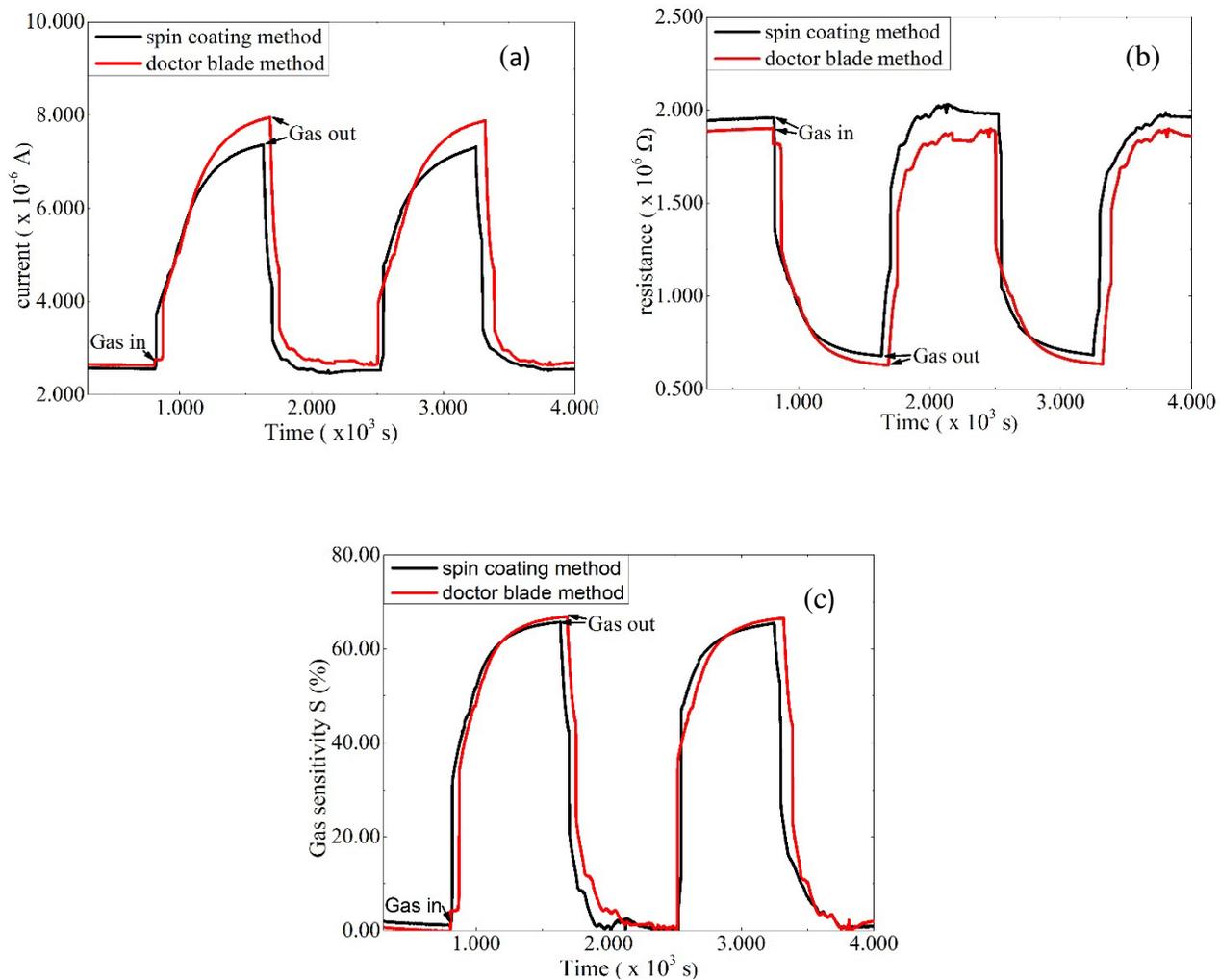

Figure 5: The graphical representations of (a) current (b) resistance (c) gas sensitivity for $CO_2$ gas sensors prepared using the doctor blade and the spin coating methods.

| $\alpha$–Fe$_2$O$_3$-PEG gas sensor | Response time (s) | Recovery time (s) | Gas sensitivity (%) |
|---|---|---|---|
| Doctor blade method | 819 | 619 | 66.91 |
| Spin coating method | 746 | 537 | 65.76 |

Table 2: Gas sensitivity, response and recovery times of $\alpha$–Fe$_2$O$_3$ films with PEG binder prepared by the doctor blade and the spin coating techniques.

## 4. Conclusion:

After heating iron acetate nanoparticles at 500 $^0$C for 2 hours in air, iron acetate powder was converted to iron oxide powder. According to XRD patterns, the single phase $\alpha$–Fe$_2$O$_3$ could be crystallized in the thin film form after annealing the sample at 500 $^0$C for 1 hour in air. The gas sensitivity of the $\alpha$–Fe$_2$O$_3$ thin films with the PEG binder synthesized by the doctor blade method is slightly higher than that of the $\alpha$–Fe$_2$O$_3$ thin films with the PEG binder fabricated by the spin coating technique. The gas sensitivities of the films synthesized by the doctor blade and the spin coating methods are 66.91% and 65.76% in 1000ppm of CO$_2$ gas, respectively. The gas sensitivity of $\alpha$–Fe$_2$O$_3$ thin films in 1000ppm of CO$_2$ can be improved by adding the PEG binder. According to UV-Visible spectrums, the optical band gaps of $\alpha$–Fe$_2$O$_3$ samples with and without binder are 1.92 and 1.98 eV, respectively. The decrease of the optical band gap and the activation energy is the reason for the improvement of the gas sensitivity. Response time slightly decreased from 821 to 819 s after adding the binder. In addition, recovery time slightly decreased from 624 to 619 s with the addition of the PEG binder. FTIR peaks observed at 456-468 cm$^{-1}$ and 549-560 cm$^{-1}$ by confirm the formation of $\alpha$–Fe$_2$O$_3$ in the thin film. The gas sensitivity of $\alpha$–Fe$_2$O$_3$ increased from 51.04% to 66.48% after adding the PEG binder.

**References:**

1. Jason S. Corneille, Jian-Wei He and D. Wayne Goodman (1995). Preparation and characterization of ultra-thin iron oxide films on a Mo(100) surface. Surface Science **338(1)**, 211-224.

2. Xianhui Zhao, Changhong Li, Qiuping Liu, Yandong Duan, Junjing He, Su Liu, Hai Wang and Song Liang (2013). Study on the preparation and properties of colored iron oxide thin films. Journal of Physics: Conference Series **419**, 012033.

3. Bonamali Pal and Maheshwar Sharon (2000). Preparation of iron oxide thin film by metal organic deposition from Fe (III)-acetylacetonate: a study of photocatalytic properties. Thin Solid Films **379** (1-2), 83-88.

4. Yingguo Peng, Chando Park and David E. Laughlin (2003). $Fe_3O_4$ thin films sputter deposited from iron oxide targets. Journal Applied Physics **93(10)**, 7957-7959.

5. C.X. Kronawitter, S.S. Mao and B.R. Antoun (2011). Doped, porous iron oxide films and their optical functions and anodic photocurrents for solar water splitting. Applied Physics Letters **98(9)**, 092108.

6. Karel Siroky, Jana Jiresova and Lubomir Hudec (1994). Iron oxide thin film gas sensor. Thin Solid Films **245 (1-2)**, 211-214.

7. J. Peng and C.C. Chai (1993). A study of the sensing characteristics of $Fe_2O_3$ gas sensing thin film. Sensors and Actuators B: Chemical **14 (1-3)**, 591-593.

8. V.V. Malyshev, A.V. Eryshkin, E.A. Koltypin, A.E. Varfolomeev and A.A. Vasiliev (1994). Gas sensitivity of semiconductor $Fe_2O_3$ based thick-film sensors to $CH_4$, $H_2$, $NH_3$. Sensors and Actuators B: Chemical **19**, 434-436.

9. G. Fan, Y. Wang, M. Hu, Z. Luo, K. Zhang and G. Li (2012). Template free synthesis of hollow ball-like nano-$Fe_2O_3$ and its application to the detection of dimethyl methylphosphonate at room temperature. Sensors **12(4)**, 4594-4604.

10. Y. Yang. H. Ma, J. Zhuang and X. Wang (2011). Morphology controlled synthesis of hematite nanocrystals and their facet effects on gas sensing properties. Inorganic chemistry **50**, 10143-10151.

11. N.V. Long, Y. Yang, M. Yuasa, C.M. Thi, Y. Cao, T. Nann and M. Nogami (2014). Gas-sensing properties of p-type α–$Fe_2O_3$ polyhedral particles synthesized via a modified polyol

method. RSC Adv **4**: 8250-8255.

12. N.V. Long, Y. Yang, C.M. Thi, Y. Cao and M. Nogami (2014). Ultra-high stability and durability of α–$Fe_2O_3$ oxide micro- and nano structures with discovery of new 3D structural formation of grain and boundary. Colloids and Surfaces A: Physicochemical and Engineering Aspects **456**. 184-194.

13. P. Samarasekara (2009). Hydrogen and Methane Gas Sensors Synthesis of Multi-Walled Carbon Nanotubes. Chinese Journal of Physics **47(3)**, 361-369.

14. P. Samarasekara (2010). Characterization of Low Cost p-$Cu_2$O/n-CuO Junction. Georgian Electronic Scientific Journals: Physics **2(4)**, 3-8.

15. P. Samarasekara and N.U.S. Yapa (2007). Effect of sputtering conditions on the gas sensitivity of Copper Oxide thin films. Sri Lankan Journal of Physics **8**, 21-27.

16. P. Samarasekara, A.G.K. Nisantha and A.S. Disanayake (2002). High Photo-Voltage Zinc Oxide Thin Films Deposited by DC Sputtering. Chinese Journal of Physics **40(2)**, 196-199.

17. K. Tennakone, S.W.M.S. Wickramanayake, P. Samarasekara and, C.A.N. Fernando (1987). Doping of Semiconductor Particles with Salts. Physica Status Solidi (a)**104**, K57-K60.

18. P. Samarasekara and Udara Saparamadu (2012). Investigation of Spin Reorientation in Nickel Ferrite Films. Georgian electronic scientific journals: Physics **1(7),** 15-20.

19. P. Samarasekara and N.H.P.M. Gunawardhane (2011). Explanation of easy axis orientation of ferromagnetic films using Heisenberg Hamiltonian. Georgian electronic scientific journals: Physics **2(6),** 62-69.

20. P. Samarasekara (2008). Influence of third order perturbation on Heisenberg Hamiltonian of thick ferromagnetic films. Electronic Journal of Theoretical Physics **5(17),** 227-236.

21. P. Samarasekara and Udara Saparamadu (2013). In plane oriented Strontium ferrite thin films described by spin reorientation. Research & Reviews: Journal of Physics-STM journals **2(2),** 12-16.

22. P. Samarasekara and Udara Saparamadu (2013). Easy axis orientation of Barium hexa-ferrite films as explained by spin reorientation. Georgian electronic scientific journals: Physics **1(9)**, 10-15.

23. P. T. Moseley, J. Norris and De Williams (1991). Techniques and mechanisms in gas

sensing. The Adam higher series on sensors publishers.
24. S. Chopra, K. McGuire, N. Gothard and A.M. Rao (2003). Selective gas detection using a carbon nanotube sensor. Applied Physics Letters **83**, 2280-2282.
25. L. Li, Z. Tong, L. ShouChun, W. LianYuan and T. YunXia (2009). Microstructure sensors based on ZnO microcrystals with contact controlled ethanol sensing. Chinese Science Bulletin **54**, 4371-4375.
26. W. XiuZhi (2012). Improved ethanol, acetone and $H_2$ sensing performances of micro sensors based on loose ZnO nanofibers. Chinese Science Bulletin **57**, 4653-4658.
27. N. Yamazoe (1991). New approaches for improving semiconductor gas sensors. Sensors and Actuators B **5**, 7-19.
28. G.F. Fine, L.M. Cavanagh, A. Afonja and R. Binions (2010). Metal oxide semiconductor gas sensors in environmental monitoring. Sensors **10**, 5469-5502.
29. C.R. Michel, A.H. Martinez, F.H. Villalpando and J.P. Moran-Lazaro (2009). Carbon dioxide gas sensing behavior in nanostructured $GdCoO_3$ prepared by a solution polymerization method. Journal of Alloys and Compounds **484**, 605-611.
30. C.V. G. Reddy, S.V. Manorama and V.J. Rao (2000). Preparation and characterization of ferrites as gas sensor Materials. Journal of Materials Science Letters **19**, 775-778.
31. S.W. Hwang, A. Umar, G.N. Dar, S.H. Kim and R.I. Badran (2014). Synthesis and characterization of iron oxide nanoparticles for phenyl hydrazine sensor applications. Sensor Letters **12**, 97-101.